\documentclass[conference,letter]{IEEEtran}

\usepackage{graphicx}  % Written by David Carlisle and Sebastian Rahtz
\usepackage{amsmath,amssymb}    % From the American Mathematical Society
\usepackage{fancyhdr}

\newcommand{\Hi} {{\cal H}^\infty}
\newcommand{\ssp}{{\cal S}_\infty(P)}
\newcommand{\bea}{\begin{eqnarray}}
\newcommand{\eea}{\end{eqnarray}}
\newcommand{\be}{\begin{equation}}
\newcommand{\ee}{\end{equation}}
\newcommand{\bd}{\begin{displaymath}}
\newcommand{\ed}{\end{displaymath}}
\newcommand{\rhp}{\mathbb{C}_{+}}
\newcommand{\R}{\mathbb{R}}

\newtheorem{definition}{Definition}[section]
\newtheorem{corollary}{Corollary}[section]
\newtheorem{lemma}{Lemma}[section]

\begin{document}

% paper title
\title{Sensitivity Minimization by Stable Controllers for a Class of Unstable
Time-Delay Systems\thanks{$^*$ This work was supported in part by
the European Commission (contract no. MIRG-CT-2004-006666) and by
T\"UB\.{I}TAK (grant no. EEEAG-105E065).}}

% authors
\author{\authorblockN{Suat G\"um\"u\c{s}soy}
\authorblockA{was with Dept. of Electrical and Computer Eng.,\\
Ohio State University, Columbus, OH 43210, U.S.A. \\
current affiliation: MIKES Inc., Akyurt, \\
Ankara, TR-06750, Turkey\\
e-mail: {\sl suat.gumussoy@mikes.com.tr}} \and
\authorblockN{Hitay \"Ozbay}
\authorblockA{Dept. of Electrical and
Electronics Eng.,\\ Bilkent University, Bilkent, Ankara TR-06800,
Turkey, \\on leave from Dept. of Electrical and Computer Eng.,\\
Ohio State University, Columbus, OH 43210, U.S.A. \\
e-mail: {\sl hitay@bilkent.edu.tr}}}

% make the title area
\maketitle

\thispagestyle{fancy} \fancyhead{} \lhead{}
\lfoot{1--4244--0342--1/06/\$20.00~\copyright~2006 IEEE}
\cfoot{}\rfoot{ICARCV 2006}
\renewcommand{\headrulewidth}{0pt}
\renewcommand{\footrulewidth}{0pt}

% abstract
\begin{abstract}
In this paper sensitivity minimization problem is considered for a
class of unstable time delay systems. Our goal is to find a stable
controller stabilizing the feedback system and giving rise to
smallest $\Hi$ norm for the sensitivity function.  This problem
has been solved by Ganesh and Pearson (1986) for finite
dimensional plants using Nevanlinna-Pick interpolation. We extend
their technique to include possibly unstable time delay systems.
Moreover, we illustrate suboptimal solutions, and their robust
implementation.

% keywords
\end{abstract}

\IEEEpeerreviewmaketitle

% introduction
\section{Introduction}
% no \PARstart
\setcounter{equation}{0} In feedback control applications, sometimes
it is desirable to have a stable controller which internally
stabilizes the closed-loop. There are many practical reasons why we
want the controller itself to be stable, \cite{V-CSS-85}. A
necessary and sufficient condition for the existence of a stable
controller stabilizing the feedback system for a given plant is the
parity interlacing property, \cite{YBL-AUT-74}. Design of such
controllers is known as strong stabilization problem and several
methods are available for its solution for MIMO or SISO finite
dimensional plants, [2-5,9,11,12,14-17,22,23]
%\cite{CZ-TAC-01,CZ-TAC-03,CC-TAC-01,CWL-CDC-03,GO05,IOS-TAC-93,JJA-ACC-90,LS-SCL-02,Petersen-ACC06,SGP-SCL-97,SS-ACC-85,ZO-TAC-99,ZO-AUT-00},
as well as different classes of SISO time delay systems,
\cite{GO-MTNS-04,Suyama91}, under $\Hi$, ${\cal H}_2$ or other
optimization constraints. Notably, the design methods in
\cite{B-CDC-96,GP-ACC-86} give optimal stable $\Hi$ controllers for
finite dimensional SISO plants as a solution to weighted sensitivity
minimization problem, other methods provide sufficient conditions to
find stable $\Hi$ controllers.

In this paper, the method of \cite{GP-ACC-86} is generalized for a
class of time-delay systems. The plants we consider may have
infinitely many right half plane poles. Optimal and suboptimal
stable $\Hi$ controllers are obtained for the weighted sensitivity
minimization problem using the Nevanlinna-Pick interpolation.

In section \ref{section:probdef}, the control problem is defined and
the structure of the plant is given. In section
\ref{section:pfacTdelay} we summarize our earlier results on the
necessary and sufficient conditions to write the plant in the given
structure for a class of possibly unstable time-delay systems. Main
results are given in section \ref{section:optdesign}. An example can
be found in section \ref{section:example}, and concluding remarks
are made in last section.

\section{Problem Definition} \label{section:probdef}

Given a single-input-single-output linear time invariant plant $P$,
sensitivity function of the feedback system is defined as
$S:=(1+PC)^{-1}$, where $C$ is the controller to be designed. We say
that the feedback system is stable if $S,PS,CS$ are stable transfer
functions (i.e. they are in  $\Hi$). Moreover, if a stable
controller, $C\in\Hi$, stabilizes the feedback system, then $C$ is
said to be strongly stabilizing, \cite{V-CSS-85}. For a given plant
$P$, the set of all strongly stabilizing controllers is denoted by
$\ssp$.

For a given minimum phase function $W(s)$, the problem of weighted
sensitivity minimization by stable controller (WSMSC) is to find
\bea \label{eq:wsm}
\gamma_s&=& \inf_{C\in\ssp} \|W(1+PC)^{-1} \|_\infty, \\
\label{eq:wsms} &=&\|W(1+PC_{\gamma_s})^{-1} \|_\infty \eea where
$\gamma_s$ is the minimum $\Hi$ cost for WSMSC and $C_{\gamma_s}\in
\ssp$ is the corresponding optimal strongly stabilizing controller.

We assume that the transfer function of the plant can be factored as
\be \label{plant:P} P(s)=\frac{m_n(s)}{m_d(s)}N_o(s) \ee where
$m_d$, $m_n$ are inner (all-pass) functions, $m_n$ is finite
dimensional and $m_d$ is infinite dimensional; $N_o$ is outer
(minimum phase) and possibly infinite dimensional.

In section \ref{section:optdesign}, we will obtain the optimal
controller $C_{\gamma_s}\in \ssp$ for the WSMSC  problem, where the
plant $P$ admits a factorization of the form (\ref{plant:P}). But
first, in the next section, we shall illustrate how this
factorization can be done for a class of possibly unstable systems
with time delays.

\section{Plant Factorization for Time Delay Systems}
\label{section:pfacTdelay}

In this section, we summarize some preliminary results from
\cite{GO-IFAC-06} on the factorization of SISO time-delay systems in
the form (\ref{plant:P}).

The plants we consider in this paper are assumed to be in the form
\be \label{eq:delayplant} P(s)=\frac{R(s)}{T(s)}=\frac{\sum_{i=1}^n
R_i(s)e^{-h_i s}}{\sum_{j=1}^m T_j(s)e^{-\tau_j s}} \ee where $R_i$
and $T_j$ are finite dimensional, stable, proper transfer functions,
and time delays $h_i$, $\tau_j$ are assumed to be positive rational
numbers, with $0\leq h_1<\ldots<h_n$ and $0\leq
\tau_1<\ldots<\tau_m$.
\begin{definition} \label{def:TDS}
Consider $R(s)=\sum_{i=1}^n R_i(s)e^{-h_i s}$ as defined above. Let
$d_i$ be the relative degree of $R_i(s)$. Then,
\begin{enumerate}
\item if $d_1<\max{\{d_2,\ldots,d_n\}}$,  $R(s)$ is called as
retarded-type time-delay system (RTDS),
\item if $d_1=\max{\{d_2,\ldots,d_n\}}$, $R(s)$ is called as neutral-type
time-delay system (NTDS),
\item if $d_1>\max{\{d_2,\ldots,d_n\}}$,
$R(s)$ is called as advanced-type time-delay system (ATDS).
\end{enumerate}
\end{definition}
The following lemma gives a necessary and sufficient condition when
a NTDS has finitely many unstable zeros.
\begin{lemma} (\cite{GO-IFAC-06}) \label{lemma:finitezeros}
Assume that $R(s)$ is a NTDS with no imaginary axis zeros and poles,
then  the system, $R$, has finitely many unstable zeros if and only
if all the  roots of the polynomial, $\varphi(r)=1+\sum_{i=2}^n
\xi_i r^{\tilde{h}_i-\tilde{h}_1}$
 has magnitude greater than $1$ where
\bea
\nonumber  \xi_i&=&\lim_{\omega\rightarrow\infty}R_i(j\omega)R_1^{-1}(j\omega) \quad \forall \;i=2,\dots,n, \\
\nonumber h_i&=&\frac{\tilde{h}_i}{N}, \quad N, \tilde{h}_i \in
Z_{+}, \; \forall \;i=1,\dots,n. \eea
\end{lemma}

By the following corollary, all SISO time-delay systems with
finitely many unstable zeros are obtained.
\begin{corollary} (\cite{GO-IFAC-06}) \label{cor:Fsystem}
The time-delay system $R$ has finitely many unstable zeros if and
only if $R$ is a RTDS or $R$ is a NTDS satisfying Lemma
\ref{lemma:finitezeros}. Time-delay systems with finitely many
unstable zeros are defined as $F$-systems.
\end{corollary}

We define the conjugate of $R(s)=\sum_{i=1}^n R_i(s)e^{-h_i s}$ as
$\bar{R}(s):=e^{-h_ns}R(-s)M_C(s)$ where $M_C$ is inner, finite
dimensional whose poles are poles of $R$. The time-delay system
$\bar{R}$ has finitely many unstable zeros if and only if $R$ is a
ATDS or Lemma~\ref{lemma:finitezeros} is satisfied by $\bar{R}$. The
time-delay system $R$ whose conjugate $\bar{R}$ has finitely many
unstable zeros is defined as an $I$-system.

The class of SISO time-delay systems with factorization
(\ref{plant:P}) is given by the following lemma.
\begin{lemma} (\cite{GO-IFAC-06}) \label{lemma:Pfac}
If $R$ is an $F$ system and $T$ is an $I$ system in
(\ref{eq:delayplant}), then $P$ can be factored as (\ref{plant:P}).
If $R$ and $T$ are irreducible and have no common factors, then $P$
has factorization (\ref{plant:P}) if and only if $R$  and $T$ are
$F$ and $I$ system respectively.
\end{lemma}

In this paper, the plant $P$, defined by (\ref{eq:delayplant}), is
assumed to satisfy the following:
\begin{enumerate}
\item[A.1] $R_i$ and $T_j$ are stable, proper, finite dimensional
transfer functions. The delays, $h_i$, $\tau_j$ are rational numbers
such that $0\leq h_1<h_2<\ldots<h_n$, and
$0\leq\tau_1<\tau_2<\ldots<\tau_m$, with $h_1=\tau_1=0$.
\item[A.2] $R$ and $T$ have no imaginary axis zeros.
\item[A.3] $R$ and $T$ are $F$ and $I$ system respectively.
\end{enumerate}

Under the above conditions $P$ can be factored as in
(\ref{plant:P}), \bea \nonumber m_d&=&M_{\bar{T}}\frac{T}{\bar{T}},
\quad m_n=M_R, \quad N_o=\frac{R}{M_R}\frac{M_{\bar{T}}}{\bar{T}}.
\eea The zeros of the inner function $M_R$ are right half plane
zeros of $R$. The unstable zeros of $\bar{T}(s)$ are the same as the
zeros of the inner function $M_{\bar{T}}$. The conjugate of $T$ has
finitely many unstable zeros since $T$ is a $I$-system.

As an example, consider the following time-delay system: \bea
\label{plant:FIplant}
\nonumber \dot{x}(t)&=&-x(t)-2\dot{x}(t-2)+2x(t-2)+u(t), \\
\nonumber y(t)&=&4x(t-3)-2\dot{x}(t-2)+2x(t-2)+u(t) \\
& &\quad\quad\quad\quad\quad\quad \eea which has the transfer
function \bd P(s)=\frac{(s+1)+4e^{-3s}}{(s+1)+2(s-1)e^{-2s}}. \ed
The plant $P$ can be written in the form of (\ref{eq:delayplant}),
\bea \nonumber
P&=&\frac{R}{T}=\frac{R_1e^{-h_1s}+R_2e^{-h_2s}}{T_1e^{-\tau_1s}+
T_2e^{-\tau_2s}}, \\
\nonumber &=&\frac{1e^{-0s}+
\left(\frac{4}{s+1}\right)e^{-3s}}{1e^{-0s}+
\left(\frac{2(s-1)}{s+1}\right)e^{-2s}}. \eea Note that $P$
satisfies assumption A.1 (i.e., $h_1=\tau_1=0$) and A.2 since it has
no imaginary axis zeros and poles. The relative degree of $R_2$ is
larger than $R_1$, therefore, $R$ is a RTDS and has finitely many
unstable zeros (it is an $F$ system). The conjugate of $T$ is \bea
\nonumber \bar{T}(s)&=&e^{-2s}T(-s)\left(\frac{s-1}{s+1}\right), \\
&=&2+\left(\frac{s-1}{s+1}\right)e^{-2s}. \eea Note that $\bar{T}$
is NTDS which satisfies Lemma \ref{lemma:finitezeros}. So, $\bar{T}$
has finitely many zeros and hence $T$ is an $I$ system. Therefore,
the plant $P$ satisfies assumption A.3. It can be shown that $R$ has
two unstable zeros at $s_{R_{1,2}}=0.3125\pm0.8548j$. Also, $T$ has
infinitely many unstable poles converging to $\ln{\sqrt{2}}\pm
j(k+\frac{1}{2})\pi$ as $k\rightarrow\infty$, which shows that the
plant $P$ has finitely many unstable zeros and infinitely many
unstable poles. By the small-gain theorem, it is clear that
$\bar{T}$ has no unstable zeros.  Now  $P$
 can be written as in (\ref{plant:P}) where
 \bea \label{factorizedplant}
 \nonumber m_d(s)&=&\frac{T(s)}{\bar{T}(s)}, \\
 \nonumber m_n(s)&=&M_R(s)=\frac{s^2-0.6250s+0.8283}{s^2+0.6250s+0.8283}, \\
 N_o(s)&=&\frac{R(s)}{M_R(s)}\frac{1}{\bar{T}(s)}.
 \eea
Note that $M_R$ is an inner function and all its zeros are unstable
zeros of $R$. Since $\bar{T}$ has no unstable zeros, $M_{\bar{T}}$
is equal to one.

In the next section, stable $\Hi$ controllers are obtained for
plants in the form (\ref{plant:P}).

\section{Stable $\Hi$ Controller Design} \label{section:optdesign}

In this section, the results of \cite{GP-ACC-86} are extended for
plants with infinitely many unstable modes. The internal stability
problem of closed-loop system can be reduced to interpolation
problem on the sensitivity function \cite{YBL-AUT-74}. This
reduction is valid also for plants with infinitely many unstable
poles and zeros. Assume that $P(s)=\frac{m_n(s)}{m_d(s)}N_o(s)$ is
as defined above with finite dimensional inner $m_n$, infinite
dimensional inner $m_d$, outer $N_o$. Note that the plant has
finitely many unstable zeros and may have infinitely unstable poles.
Let the weighting function, $W$, be minimum phase, then the
closed-loop system is internally stable if and only if there exists
$S_W\in\Hi$, $S_W=W(1+PC)^{-1}$ satisfying \be \label{eq:SW} S_W(s)=
m_d(s)F_\gamma(s) \ee where $F_\gamma\in \Hi$, and
\be\label{def:int_cond_Fg} m_d(s_i)F_\gamma(s_i)=W(s_i), \ee for all
zeros of $m_n(s)$, $s_i\in\rhp$, $i=1\dots ,N$. Moreover, $\|
S_W\|_\infty =\| F_\gamma \|_\infty$. Optimal weighted sensitivity
is the one which corresponds to an $F_\gamma$ whose $\Hi$ norm is
the smallest among all stable functions satisfying
(\ref{def:int_cond_Fg}).

When the controller in the weighted sensitivity minimization problem
defined above is restricted to be stable, then we must have \bea
\nonumber C_{\gamma}&=&\frac{W-S_W}{S_WP}=\frac{(W- \gamma m_d F
)N_o^{-1}}{\gamma m_d F P}, \\
\nonumber &=&\frac{(W-\gamma m_d F)N_o^{-1}}{\gamma m_n F}\in\Hi
\eea where  $F \in \Hi$ and $F^{-1}\in \Hi$ with $\|F \|_\infty\le
1$ and it satisfies the interpolation conditions \be
\label{eq:interpconds} F(s_i)=\frac{W(s_i)}{\gamma
m_d(s_i)}=\frac{\omega_i}{\gamma},\quad i=1,\ldots,N \ee for the
smallest possible $\gamma > 0$. Conversely, if there exists such an
$F$, then optimal stable $\Hi$ controller $C_{\gamma_s}$ for WSMSC
problem (\ref{eq:wsm}) can be obtained from $S_W$. The optimal $\Hi$
cost for (\ref{eq:wsms}) is $\gamma_s$, which is the smallest
$\gamma$ value for which a unit $F\in \Hi$ satisfying
(\ref{eq:interpconds}) can be found. (We say that a function $F\in
\Hi$ is a unit if $F^{-1}\in\Hi$ and $\|F\|_\infty\leq 1$). Note
that the above transformation reduces the WSMSC problem for plants
with infinitely unstable modes into an interpolation problem, by a
unit in $\Hi$, with finitely many interpolation conditions.

The solution of the interpolation problem with unit is given in
\cite{GP-ACC-86} using the Nevanlinna-Pick approach,
%\cite{FOT96,KN77,ZO_NevPick}
[6,13,21], as follows. Define \be
\label{def:G(s)} G(s)=-\ln{F(s)} ~~~~F(s)=e^{-G(s)}. \ee Now, we
want to find an analytic function $G~:~\rhp \rightarrow \rhp$ such
that \bd G(s_i)=-\ln{\omega_i}+\ln{\gamma}-j2\pi m_i=:\nu_i,\quad
i=1,\ldots,N \ed where $m_i$ is a free integer due to non-uniqueness
of the complex logarithm. Note that when $\|F\|_\infty \le 1$ the
function $G$ has a positive real part hence it maps $\rhp$ into
$\rhp$. Now if the extended right-half plane is transformed onto the
closed unit disc in the complex plane by one-to-one conformal
mapping $z=\phi(s)$, then the transformed interpolation conditions
are \be \label{eq:interpcondz} f(z_i)=\frac{\omega_i}{\gamma},\quad
i=1,\ldots,N \ee where $z_i=\phi(s_i)$ and $f(z)=F(\phi^{-1}(z))$.
The transformed interpolation problem is to find a unit with
$\|f\|_\infty\leq 1$ such that interpolation conditions
(\ref{eq:interpcondz}) are satisfied. By the following
transformation, \be \label{trans:lnf} g(z)=-\ln{f(z)}, \ee the
interpolation problem can be written as, \bd g(z_i)=\nu_i,\quad
i=1,\ldots,N. \ed Define $\phi(\nu_i)=:\zeta_i$. If we can find an
analytic function $\tilde{g}$ mapping unit disc onto unit disc,
satisfying \bd \tilde{g}(z_i)=\zeta_i ~~~~i=1,\ldots,N \ed then the
desired $g(z)$, hence $f(z)$ and $F(s)$ can be constructed from
$g(z)=\phi^{-1}(\tilde{g}(z))$. The problem of finding such
$\tilde{g}$ is the well-known Nevanlinna-Pick problem,
%\cite{FOT96,KN77,ZO_NevPick}
[6,13,21]. The condition for the existence of an
appropriate $g$ can be given directly: there exists an analytic $g$
mapping the unit disc onto right half plane if and only if the Pick
matrix $P_{N\times N}$, \be \label{eq:pickmatrix}
P(\gamma,\{m_i\})_{i,k} =\left[\frac{2\ln{\gamma}-\ln{\omega_i}
-\ln{\bar{w}_k}+j2\pi m_{k,i}}{1-z_i\bar{z}_k}\right] \ee is
positive semi-definite, where $m_{k,i}=m_k-m_i$ are integers. In
\cite{GP-ACC-86}, it is mentioned that the possible integer sets
$\{m_i\}$ are finite and in all possible integer sets $\{m_i\}_l,\;
l=1,\ldots,r$, there exists a minimum value, $\gamma_s$, such that
$P(\gamma_s,\{m_i\}_l)\geq~0$.

\subsection{Optimal Stable $\Hi$ Controller Design Algorithm for Plants with Infinitely Unstable Modes:} \label{algorithm}
\begin{enumerate}
    \item Write the plant in the form of (\ref{plant:P}):
    \begin{enumerate}
    \item[]If the plant is a SISO time-delay system, obtain its
    transfer function and re-write it in the form of (\ref{eq:delayplant}).
    If $R$ and $T$ satisfy Assumptions A.1-A.3, do the
    factorization of the plant as (\ref{plant:P}).
    \end{enumerate}
    \item Find the zeros  $s_i\;\;i=1,\ldots,N$ of $m_n(s)$.
    \item Calculate $\omega_i$, and using a conformal mapping $\phi$
    calculate $z_i$ for $i=1,\ldots,N$.
    \item For all possible integer sets $\{m_i\}_l,\; l=1,\ldots,r$,
    find $\gamma_s$ such that the Pick matrix (\ref{eq:pickmatrix})
    is positive semi-definite.
    \item Obtain optimal interpolation function $g_{\gamma_s}(z)$
    and $f_{\gamma_s}(z)$ by transformation (\ref{trans:lnf}),
    see e.g. %\cite{FOT96,ZO_NevPick}
    [6,21].
    \item Calculate $F_{\gamma_s}(s)=f_{\gamma_s}(\phi(s))$ and
    $S_{W,\gamma_s}(s)$ using $F_{\gamma_s}$ and $\gamma_s$ in
    (\ref{eq:SW}).
    \item The optimal stable $\Hi$ controller for plants with
    infinitely unstable modes can now be written as
    \bd
    C_{\gamma_s}=\frac{W-S_{W,\gamma_s}}{S_{W,\gamma_s}P}.
    \ed
    Note that this controller achieves the optimal $\Hi$ norm
    $\gamma_s$ which is the minimum value
    for WSMSC problem.
\end{enumerate}

\subsection{Remarks:}

1) Clearly, stable $\Hi$ controller design is also applicable to
infinite dimensional plants with finitely many right half plane
poles and zeros. In this case it is possible to write the plant as
\be \label{plant:FFplant} P(s)=\frac{m_n(s)}{m_d(s)}N_o(s) \ee where
$m_n$ and $m_d$ are finite dimensional inner functions whose zeros
are $\rhp$ zeros and poles of plant $P$ respectively; $N_o$ is
outer, i.e. the minimum phase part of the plant $P$. For time-delay
systems (\ref{eq:delayplant}), this case means that $R$ and $T$ are
$F$ systems. Stable $\Hi$ controller design for plants
(\ref{plant:FFplant}) is the same as in \cite{GP-ACC-86}. The main
difference is that the term $m_d$ in $S_W$ is finite dimensional.
There are many plants with the above structure, such as,
\bea \label{plant:ssFF}
\nonumber \dot{x}(t)&=&\sum_{i=0}^{n_A}A_ix(t-h_{A,i})+B u(t), \\
y(t)&=&C x(t)+du(t)
\eea
where $ A_i \in \R^{n \times n}$,
$i=1,\ldots,n_A$ and $B,C,d$ are real valued vectors of appropriate
dimensions. The state vector has dimension is
$x(t):=[x_1(t),\ldots,x_n(t)]^T$ and the time-delays satisfy \bea
\nonumber 0&\leq&h_{A,1}< \ldots < h_{A,i} < \ldots < h_{A,n_A}.
\eea Optimal stable $\Hi$ controller can be found for the plant
(\ref{plant:ssFF}).

2) Note that optimal stable $\Hi$ controller is unique and it is not
rational.  For practical purposes, rational approximation of the
optimal controller can be done with desired error bound or a
rational controller can be searched in the set of suboptimal
controllers determined from the suboptimal solutions of the
Nevanlinna-Pick problem.

3) There are always unstable pole-zero cancellations in the
controller terms, $\frac{W-S_{W,\gamma_s}}{m_n}$ and $N_o$ from
interpolation conditions and factorization respectively. It is not
possible to directly cancel the unstable pole-zeros since the
optimal interpolating function $F$ in $S_{W,\gamma_s}$ is
irrational. If the suboptimal controllers are considered, the
interpolating function $F$ can be chosen as finite dimensional.
Exact cancellations are possible for infinite dimensional plants
(\ref{plant:FFplant}) with finite dimensional $F$ in the term
$\frac{W-S_{W,\gamma_s}}{m_n}$. If $F$ is finite dimensional and the
plant is a time-delay system with factorization (\ref{plant:P}), the
controller can be written in a form such that the controller has a
finite impulse response structure which eliminates unstable
pole-zero cancellation problem in $\frac{W-S_{W,\gamma_s}}{m_n}$ and
$N_o$, see \cite{GO-IFAC-06}. This new structure of controller makes
possible to implement the controller practically. The example shows
this structure in~\ref{section:example}.

\section{Example} \label{section:example}

Optimal stable $\Hi$ controller for WSMSC problem is designed for
time-delay plant (\ref{plant:FIplant}) with weighting function
$W(s)=\frac{1+0.1s}{s+1}$. The time-delay system
(\ref{plant:FIplant}) is put in the form of (\ref{factorizedplant}).
The zeros of plants are $s_{R_{1,2}}=0.31\pm0.85j$, and
$\omega_{1,2}=0.79\mp0.42j$. This gives the optimal $\Hi$ cost
$\gamma_s=1.07$.

The algorithm gives the optimal $\Hi$ cost for WSMSC problem, that
is the best value for any stable controller. Unfortunately, the
resulting optimal stable $\Hi$ controller has internal unstable
pole-zero cancellations. If the suboptimal case is considered, a
practical controller can be found.

\begin{figure}[h]
\begin{minipage}[t]{4cm}
\begin{center}
\includegraphics[width=4cm,]{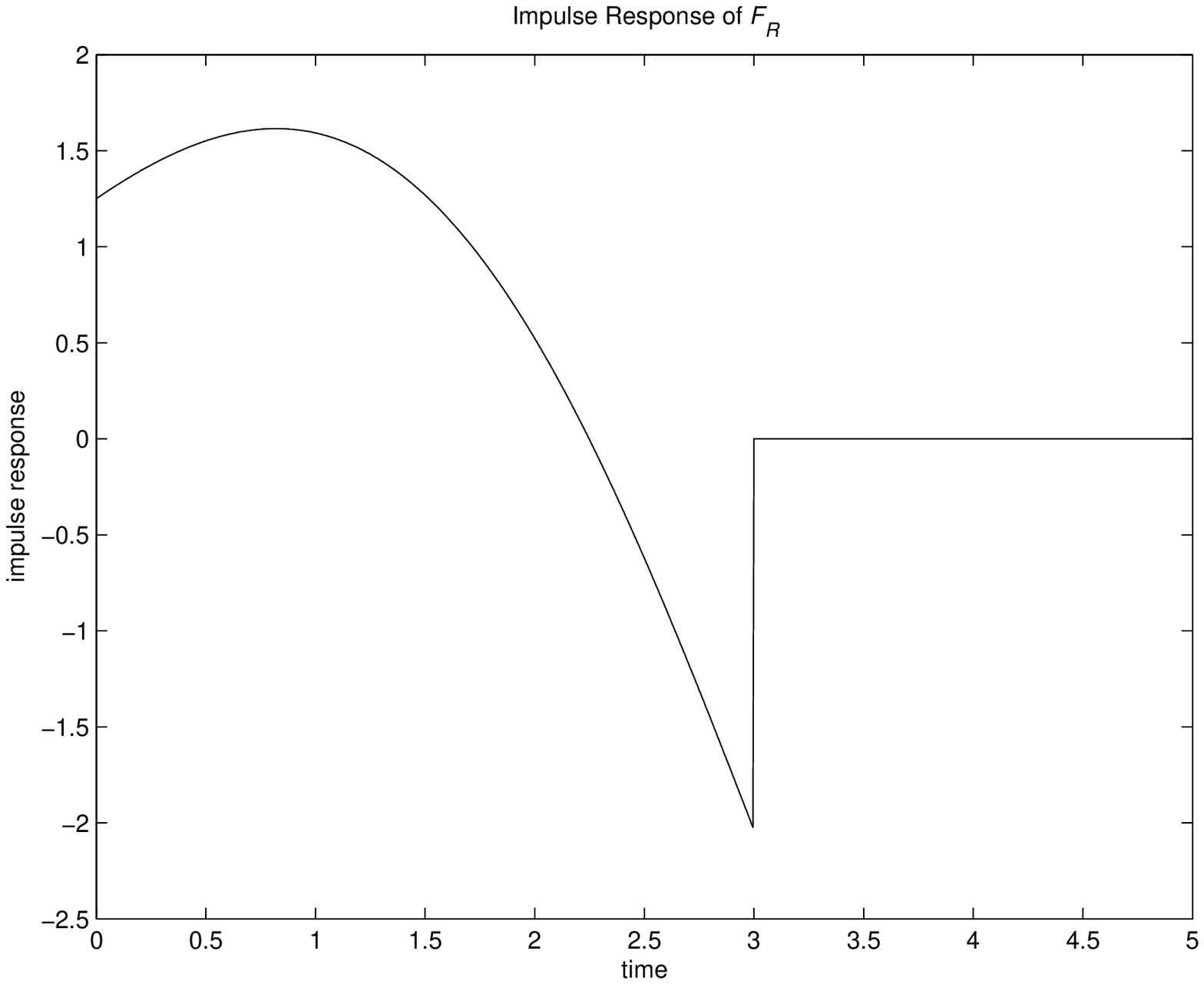}
\end{center}
\end{minipage}
\hfill
\begin{minipage}[t]{4cm}
\begin{center}
\includegraphics[width=4cm]{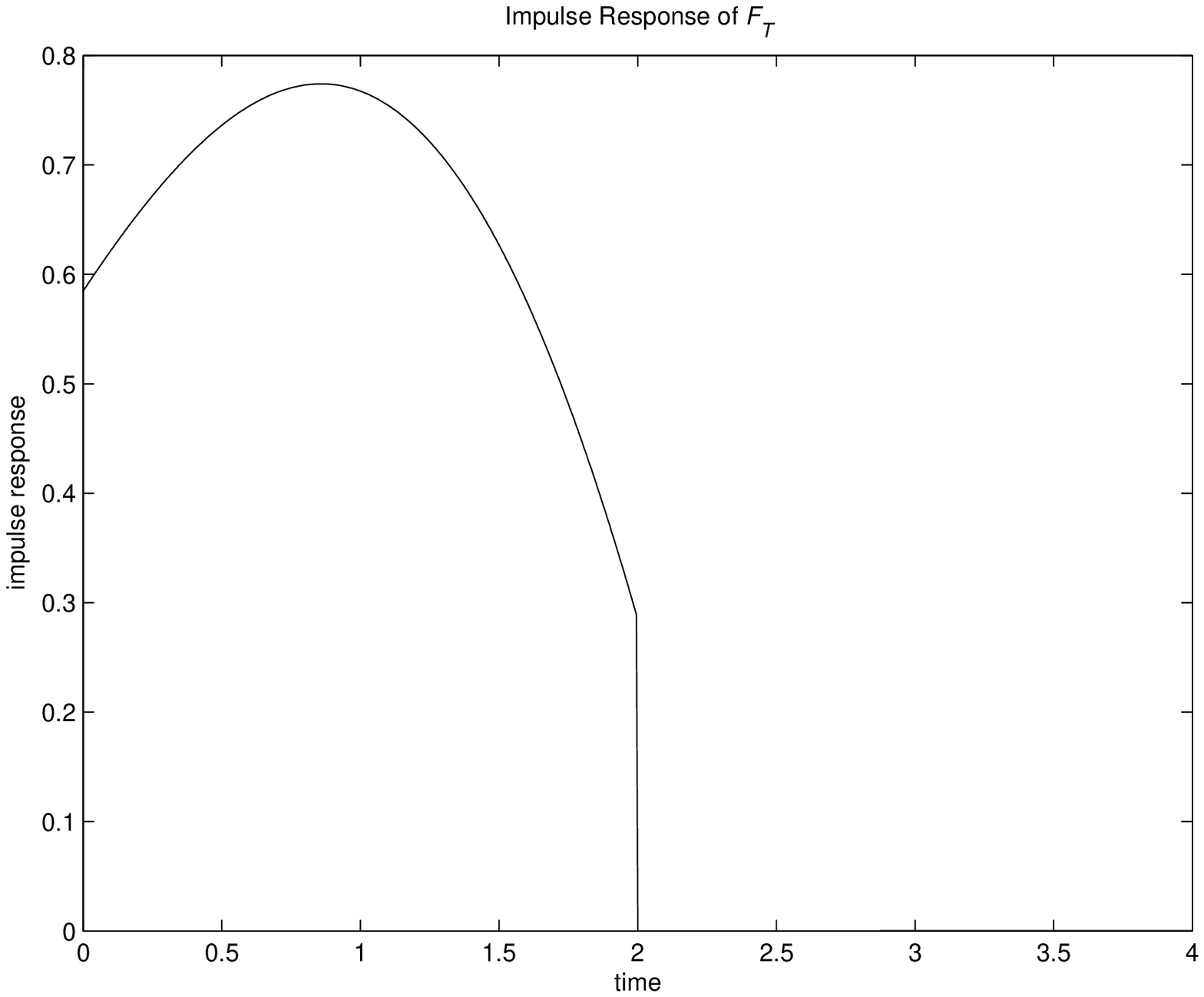}
\end{center}
\end{minipage}
\caption{\label{Fig1} Impulse Responses of $\mathcal{F}_R$ and
$\mathcal{F}_T$}
\end{figure}

Consider a suboptimal solution to WSMSC for $\gamma=1.5$ which is
larger than the optimal cost, $\gamma_s=1.07$. By a numerical search
algorithm, a finite dimensional interpolating function $F_{subopt}$
can be found as \bd F_{subopt}(s)=\frac{0.1895s+0.7308}{s+0.7310}.
\ed Note that $F_{subopt}$ is a unit with
$\|F_{subopt}\|_\infty\leq1$ and satisfies the interpolation
conditions $F_{subopt}(s_{R_{i}})=\gamma^{-1}\omega_{i}$ for
$i=1,2$. The corresponding suboptimal sensitivity function can be
obtained as $S_{W,\gamma}=\gamma m_d F_{subopt}$. The suboptimal
stable $\Hi$ controller is \bea
\nonumber C_\gamma&=&\frac{W-S_{W,\gamma}}{S_{W,\gamma}P}, \\
\nonumber
&=&\left(\frac{\gamma^{-1}WF^{-1}_{subopt}-m_d}{m_n}\right)\frac{1}{N_o}.
\eea Note that there are unstable pole-zero cancellations inside the
parenthesis in the above expression, and in $N_o$. It is clear that
when the infinite dimensional plant admits a factorization
(\ref{plant:FFplant}), exact cancellation inside the parenthesis is
possible because all the terms are finite dimensional. However, the
plant in this example has an infinite dimensional part, $m_d$, so it
is not possible to make exact cancellations in the controller.
Nevertheless, the unstable pole-zero cancellations can be avoided by
the method proposed in \cite{GO-IFAC-06} as follows: \bea \nonumber
C_\gamma&=& \left(\frac{\gamma^{-1}WF^{-1}\bar{T}-T}{m_n}\right)
\left(\frac{R}{m_n}\right)^{-1}, \\
\nonumber &=&(H_T+\mathcal{F}_T)(H_R+\mathcal{F}_R)^{-1} \eea where
$\mathcal{F}_T$ and $\mathcal{F}_R$ are finite impulse response
filters (i.e. their impulse responses are non-zero only on a finite
time interval) \bea \nonumber
\mathcal{F}_R(s)&=&\frac{1.25s+(2.04s+1.69)e^{-3s}}
{s^2-0.625s+0.828}, \\
\nonumber
\mathcal{F}_T(s)&=&\frac{0.585s+0.019-(0.285s-1.066)e^{-2s}}
{s^2-0.625s+0.828}, \eea whose denominators are determined from the
zeros of $m_n$. The impulse responses of $\mathcal{F}_T$ and
$\mathcal{F}_R$ are given in Figure \ref{Fig1}. The terms, $H_R$ and
$H_T$, are time-delay systems with no unstable pole-zero
cancellations internally.

Note that if the plant has factorization (\ref{plant:FFplant}),
$\mathcal{F}_T=0$ since $m_d$ is finite dimensional. The exact
cancellations can be made in $\frac{W-S_{W,\gamma}}{m_n}$ and $F_R$
is from unstable pole-zero cancellations inside $N_o$.

\section{Concluding Remarks}
Weighted sensitivity minimization problem by stable $\Hi$
controllers is considered for SISO infinite dimensional plants with
finitely many right half plane zeros and possibly infinitely many
right half plane poles. The optimal stable $\Hi$ controller and
corresponding optimal $\Hi$ cost are obtained from the
Nevanlinna-Pick interpolation. For this purpose the approach of
\cite{GP-ACC-86} is extended to the class of unstable time delay
systems considered here. Suboptimal controllers can be found from
all suboptimal interpolants determined by the Nevanlinna-Pick
solutions, and infinite dimensional suboptimal controllers can be
approximated by finite dimensional ones. It should be noted that
when the plant has infinitely many right half plane zeros, there
will be infinitely many interpolation conditions, and this approach
will not be applicable in such cases. Another open problem in this
area is the extension of the main results to a two-block $\Hi$
control problem, for example the mixed sensitivity minimization.

\bigskip
\noindent {\bf Acknowledgements}: This work was supported in part
by the European Commission (contract no. MIRG-CT-2004-006666) and
by T\"UB\.{I}TAK (grant nos. EEEAG-105E065 and EEEAG-105E156).

\end{document}